# Spectral Line-by-Line Pulse Shaping of an On-Chip Microresonator Frequency Comb

Fahmida Ferdous,[1] Houxun Miao,[2,3*] Daniel E. Leaird,[1] Kartik Srinivasan,[2] Jian Wang,[1,4]
Lei Chen,[2] Leo Tom Varghese,[1,4] and A. M. Weiner[1,4*]

[1] ECE Department, Purdue University, 465 Northwestern Avenue, West Lafayette, Indiana 47907, USA

[2] Center for Nanoscale Science and Technology, National Institute of Standards and Technology, 100 Bureau Dr, Gaithersburg, MD 20899, USA

[3] Maryland Nanocenter, University of Maryland, College Park, MD 20742, USA

[4] Birck Nanotechnology Center, Purdue University, 1205 West State Street, West Lafayette, Indiana 47907, USA

[*]Corresponding authors: amw@purdue.edu , houxun.miao@nist.gov

We report, for the first time to the best of our knowledge, spectral phase characterization and line-by-line pulse shaping of an optical frequency comb generated by nonlinear wave mixing in a microring resonator. Through programmable pulse shaping the comb is compressed into a train of near-transform-limited pulses of ≈ 300 fs duration (intensity full width half maximum) at 595 GHz repetition rate. An additional, simple example of optical arbitrary waveform generation is presented. The ability to characterize and then stably compress the frequency comb provides new data on the stability of the spectral phase and suggests that random relative frequency shifts due to uncorrelated variations of frequency dependent phase are at or below the 100 microHertz level.

Optical frequency combs consisting of periodic discrete spectral lines with fixed frequency positions are powerful tools for high precision frequency metrology, spectroscopy, broadband gas sensing, optical clocks, and other applications [1-7]. Frequency combs generated in mode locked lasers can be self-referenced to have both stabilized optical frequencies and repetition rates (with repetition rates below ~1 GHz in most cases) [8]. An alternative approach based on strong electro-optic phase modulation of a continuous wave (CW) laser provides higher repetition rates, up to a few tens of GHz, useful for applications such as telecommunications, but without stabilization of the optical frequency [9-11]. Optical frequency comb generators based on electro-optic phase modulation in an optical cavity are also known [12]. Recently, on-chip comb generation methods based on nonlinear optical modulation in ultrahigh quality factor (Q) monolithic microresonators have been demonstrated, where two pump photons are transformed into two sideband photons in a four wave mixing (FWM) process mediated by the Kerr nonlinearity [13-19]. The small confinement volume, high photon density, and long photon storage time in the resonator induce a very strong light-matter interaction which leads to a significant reduction in the threshold of nonlinear optical processes. The repetition rate or frequency spacing of such Kerr combs is typically hundreds of GHz. A cascade of the FWM process is believed to ensure that the frequency difference of pump and first-order sidebands is exactly transferred to all higher-order inter-sideband spacing. Thus, provided that the cavity exhibits a sufficiently equidistant mode spacing, successive FWM to higher orders intrinsically leads to the generation of a multiplicity of sidebands with equal spacing, that is, an optical frequency comb. The essential advantages of this method for comb generation are simplicity, small size, and very low power consumption.

Experiments investigating the time domain characteristics of waveforms produced by Kerr combs are at best very limited. One recent study presented autocorrelation data suggesting that different waveforms are produced every time the Kerr comb is turned on [19]. Knowledge of the phase relationship between the various spectral lines is important both from a fundamental perspective and to allow manipulation of the time domain behavior, e.g., through programmable optical pulse shaping [20]. Recently, increasing interest has arisen in line-by-line shaping, in which the individual lines of a frequency comb are resolved and independently controlled [21-25]. Also termed optical arbitrary waveform generation, pulse shaping at the individual line level offers significant opportunities for impact both in technology (e.g., telecommunications, lidar) and ultrafast optical science (e.g., coherent control and spectroscopy). Here for the first time we demonstrate spectral phase characterization of a Kerr comb generated from a microresonator, application of pulse shaping for line-by-line phase correction resulting in a 595 GHz train of near-bandwidth-limited, ≈ 300 fs pulses, as well as a first example of optical arbitrary waveform generation from such a source. An important feature of our approach is that transform-limited pulses may in principle be realized for any spectral phase signature arising in the comb generation process. Furthermore, the ability to achieve successful pulse compression provides new information on the stability of the frequency dependent phase of Kerr combs. Our results suggest that random relative frequency shifts due to uncorrelated variations of frequency dependent phase are at or below the 100 microHertz level.

We fabricated silicon nitride ring resonators for the frequency comb generation. We started with a (100) silicon wafer. A 3 µm thick silicon dioxide layer was grown in a thermal oxidation furnace. Then, a 430 nm thick silicon nitride layer was deposited using low-pressure chemical vapour deposition (LPCVD). The nitride layer was patterned with electron-beam lithography and etched through using a reactive ion etch (RIE) of $CHF_3/O_2$ to form microring resonators (40 µm radius and 2 µm width) and waveguides (1 µm width, 700 nm ring-waveguide gap) coupling light into and out of the resonators. The waveguides were linearly tapered down to a width of 100 nm at their end for low loss coupling to/from optical fibers [26]. Another 3 µm oxide layer was deposited using a low temperature oxide (LTO) furnace for the top cladding. The wafer was annealed for 3 h at 1200 °C in an ambient $N_2$ environment. A photolithography and liftoff process was used to define a metal mask for V-grooves that provide a self-aligned region where the on-chip waveguide inverse tapers are accessible by an optical fiber placed in the V-groove. After mask definition, the V-grooves were formed by RIE of the unprotected oxide and nitride layers and KOH etching of the silicon.

Fig. 1(a) shows the microscope image of one of the fabricated microring resonators with coupling waveguide. For robust and low-loss coupling of light into and out of the devices, we have developed a process for fiber pigtailing the chip, as shown in Fig. 1 (b). The silicon nitride waveguide taper of 100 nm at the end is surrounded by the cladding oxide of ≈ 10 µm in width, which provides both mechanical support and coupling enhancement. Single mode optical fibers are placed in the V-grooves and aligned with the waveguides, with coupling losses per facet of ≈ 6.4 dB and ≈ 3 dB achieved for bare fibers and lensed fibers, respectively. Epoxy is applied to the fiber V-groove interface and cured with a UV lamp to form a fiber pigtailed device. For simplicity, we use bare fibers as pigtails. The fiber pigtailing eliminates the time consuming task of free space coupling and significantly enhances the transportability of these devices.

Spectroscopy of the resonator's optical modes is performed with a swept-wavelength tunable diode laser with time-averaged linewidth of less than 5 MHz. Figure 1(c) shows the transmission spectrum of the transverse magnetic (TM) modes, which have their electric field vectors predominantly normal to the plane of the resonator. Two orders of modes are observed, where one yields deep resonances with high Qs, and the other produces more shallow resonances with lower Qs. Fig. 1 (d) shows a zoomed in spectrum for a mode at ≈ 1556.43 nm with a line width of 1.2 pm, corresponding to an optical Q of $1.3 \times 10^6$. Considering the resonance transmission contrast of the mode ($\Delta T=0.9$), the intrinsic Q is estimated to be $2.0 \times 10^6$. The average free spectral range for the series of high-Q modes is measured to be ≈ 4.7 nm.

Figure 2 shows the experimental setup and a generated spectrum (taken from an optical spectrum analyzer before the pulse shaper) by launching strong CW pumping light (we estimate 27.5 dBm to the waveguide) to a mode at around 1543 nm. Twenty six comb lines (including the pump) are generated with a comb spacing of ≈ 4.7 nm as expected given the micro-ring size. Although the generated bandwidth is narrower than previously reported for silicon nitride [14, 19], the number of lines is already sufficient for line-by-line pulse shaping experiments. Previous research [14] indicates that engineering of the waveguide design for optimized dispersion, which has not yet been performed in our devices, is critical to achieving the widest comb bandwidth. We intend to carry out such dispersion optimization for future experiments on comb generation and shaping.

The generated frequency comb is launched to a line-by-line pulse shaper, which both assists in spectral phase characterization and enables pulse compression and shaping. Line-by-line pulse shaping is implemented using a fiber-coupled Fourier-transform pulse shaper that incorporates a $2 \times 128$ pixel liquid crystal modulator (LCM) array to independently control both the intensity and phase of each spectral line [20]. The output waveform from the pulse shaper is fed to intensity autocorrelation setup through a fiber amplifier. The path from the output of the microring chip to the autocorrelator comprises $\approx$ 18 m of standard single mode fiber (SMF). Spectral phase measurement and compression are accomplished simultaneously using the method reported in [27]. Briefly, the phase is corrected by adding one comb line at a time and adjusting the phase of the new frequency component to maximize the second harmonic generation (SHG) signal from the autocorrelator at zero delay. To optimize the SHG signal, the phase of the new frequency component is varied from 0 to $2\pi$ in steps of $\pi/12$. The SHG signal should be maximized when the phase applied is equal but opposite to the original phase of that frequency component. Once the SHG is optimized, the pulses are compressed close to the bandwidth limit, and the opposite of the phase applied on the pulse shaper gives an estimate of the original spectral phase after the comb has propagated to the autocorrelator. In [28], the phase measurement by this SHG optimization method is compared with another independent method based on spectral shearing interferometry. The differences between the two measurements were comparable to the $\pi/12$ step size of the SHG optimization method. This provides an estimate of both the precision and the accuracy of our phase measurement method. After knowing the spectral phases, we can apply user defined additional phases and amplitudes to the comb lines for optical arbitrary waveform generation.

We select 10 comb lines (limited by the bandwidth of the pulse shaper) to perform the line-by-line pulse shaping experiments. Fig. 3(a) shows the spectrum of the comb source after the pulse shaper. In addition to phase compensation, here we have programmed the LCM to flatten the power spectrum, which leads to better pulse quality. Normally in our experiments, only the strong CW line at the pump frequency and a moderately strong line nearest to the pump at lower wavelength (see Fig. 2) are significantly attenuated. We then find that the SHG signal is maximized when we apply the spectral phase profile of Fig. 3(b). Fig. 3(c) shows the measured autocorrelation traces of the spectrally flattened comb both before and after spectral phase correction. The signal indicates only a weak intensity modulation without phase correction, while a clear pulse-like signature is present after correction. The intensity profile of the compressed pulse, calculated based on the spectrum shown in Fig. 3(a) and assuming a flat spectral phase, has a duration of 270 fs FWHM. The effect of spectral phase measurement precision is evaluated through simulations in which random spectral phases in the range 0 to $\pi/12$ are applied to the comb lines. This results only in minimal variations (below 1 fs) in the predicted pulse duration. The intensity autocorrelation function of the calculated compressed pulses is also computed and, as shown in Fig. 3(c), is in excellent agreement with the experimental trace. The widths of experimental and computed autocorrelation traces are 411 fs and 394 fs, respectively, which differ only by 4%. From this we take the uncertainty in our 270 fs pulse duration estimate as ±4%.

We noticed that the performance of the resonator degraded after exposing to the air for a long time (a few months). The number of generated comb lines reduced to 9. We performed line-by-line pulse shaping for the generated frequency comb and again we were able to compress to nearly bandwidth-limited pulses. Fig. 4 shows the spectrum after equalization by the pulse shaper, the applied spectral phase, and the corresponding autocorrelation traces. The transformation of the signal from nearly unmodulated in time (before phase correction) to pulse-like is even more dramatic than in the previous example. Again experimental and calculated autocorrelations are in good agreement, with FWHM autocorrelation widths equal to 460 fs and 427 fs, respectively. The duration of the calculated intensity profile, obtained via the procedure explained above is 312 fs FWHM, with uncertainty estimated at ±7% based on the autocorrelation results.

As a preliminary example of arbitrary waveform generation with the generated frequency comb, we program the pulse shaper to apply a $\pi$-step function to the spectrum of the compressed pulse in Fig. 4(b)-(c). Fig. 4(d) shows the results. The $\pi$ step occurs at pixel number 64 (corresponding to 1550 nm in wavelength). Application of a $\pi$ phase step onto half of the spectrum is known to split an original pulse into an electric field waveform that is antisymmetric in time, sometimes termed an odd pulse [29]. This corresponds to a doubly peaked intensity waveform with triply peaked intensity autocorrelation. The resulting triplet is clearly visible as shown in Fig. 4(d) and is in good agreement with the autocorrelation that is computed based on the spectrum shown in Fig. 4(a) and a spectral phase function that is flat except for a $\pi$ step centered at 1550 nm. This result constitutes a clear example of line-by-line pulse shaping.

The ability to perform such pulse compression and shaping provides clear evidence of phase coherence across the comb spectrum. In Fig. 4(e) we show compressed pulse autocorrelation measurements taken at three times over a one hour time interval. Here the spectral phase profile applied by the pulse shaper is the same for all measurements. Clearly the compression results remain similar over the time period, which means that the relative phases of the comb lines must remain approximately fixed; uncorrelated variation in relative spectral phase must conservatively remain substantially below π. Generally phase and frequency are related through the expression $\Delta f = \Delta \phi / (2\pi \Delta t)$. Taking phase variations $\Delta\varphi < 0.7\pi$ and observation time $\Delta t = 3600$ s, we estimate that random relative frequency shifts Δf due to uncorrelated variations of frequency dependent phase are of the order $10^{-4}$ Hz or below.

In summary, we have demonstrated line-by-line optical pulse shaping on a frequency comb generated from a SiN microring resonator. Nearly bandwidth-limited optical pulses were achieved after spectral phase correction. An example of arbitrary wave generation was demonstrated. The ability to controllably compress and reshape combs generated through nonlinear wave mixing in microresonators provides new evidence of phase coherence across the comb spectrum. Furthermore, in future investigations the ability to extract the phase of individual lines may furnish clues into the physics of the comb generation process.

We gratefully acknowledge Vladimir Aksyuk, Minghao Qi, Chris Long, and Victor Torres Company for their comments and suggestions and thank the staff of the CNST Nanofab, especially Richard Kasica, for assistance with E-beam lithography. This project was supported in part by the National Science Foundation under grant ECCS-0925759 and by the Naval Postgraduate School under grant N00244-09-1-0068 under the National Security Science and Engineering Faculty Fellowship program. Any opinion, findings, and conclusions or recommendations expressed in this publication are those of the authors and do not necessarily reflect the views of the sponsors.

Fig.1 (a) Microscope image of a 40 µm radius silicon nitride micro-ring with coupling region. (b) Image of a fiber pigtail. (c) Transmission spectrum of the microring resonator. (d) Zoomed in spectrum of an optical mode with 1.2 pm linewidth.

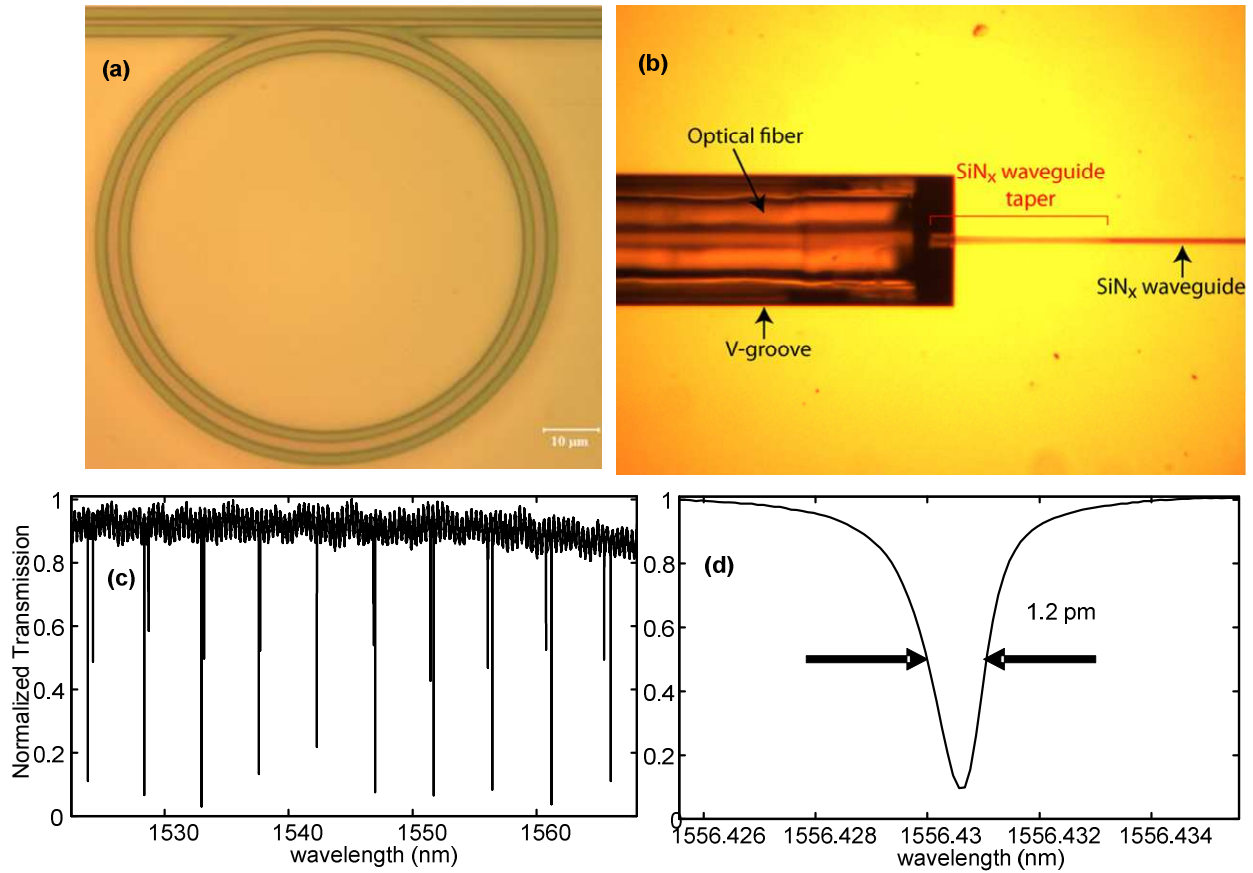

Fig.2 (a) Scheme of the experimental setup for line-by-line spectral shaping of frequency comb from silicon nitride microring. CW: tunable continuous-wave laser; μ ring: SiN microring; OSA: optical spectrum analyser; AC: autocorrelator. (b) Spectrum of a generated optical frequency comb.

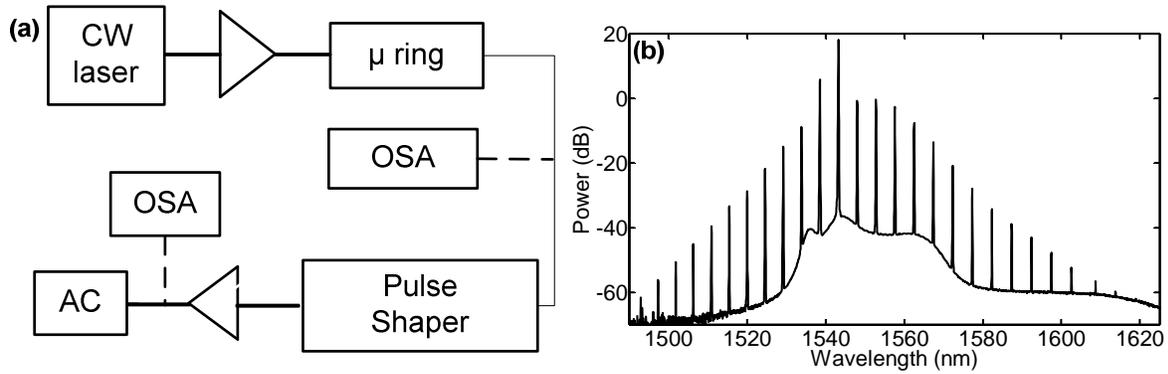

Fig. 3 (a) Spectrum of the generated comb after the pulse shaper. (b) The phase applied to the liquid crystal modulator (LCM) pixels of the pulse shaper for optimum SHG. (c) Auto-correlation traces. Red line is the compressed pulse after phase correction, dark blue line is the uncompressed pulse and black line is the calculated trace by taking the spectrum shown in Fig. 3(a) and assuming flat spectral phase.

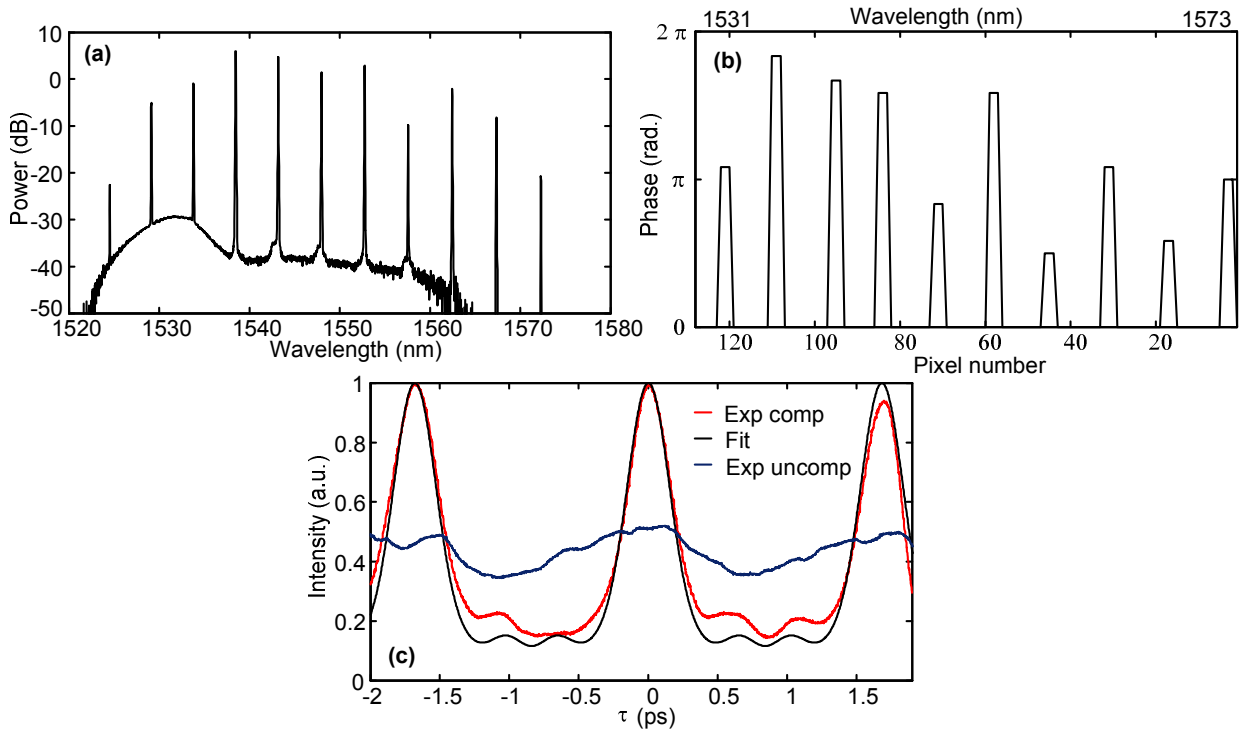

Fig.4 (a) Spectrum of the generated comb after the pulse shaper. (b) The phase applied to the LCM pixels of the pulse shaper for optimum SHG. (c) Autocorrelation traces. Here red line is the compressed pulse, dark blue line is the uncompressed pulse, and black line trace is the calculated trace by taking the spectrum shown in fig 4(a) and assuming flat spectral phase. (d) The odd pulse: applied phase same as Fig. 4(b), but with additional $\pi$ phase added for pixels 1-64 (wavelengths longer than 1550 nm). Red line: experimental autocorrelation; black line: autocorrelation calculated using the spectrum of Fig. 4(a), with a $\pi$ step centered at 1550 nm in the spectral phase. (e) Normalized intensity autocorrelation traces for compressed pulses, measured respectively at 0, 14, and 62 minutes after spectral phase characterization.

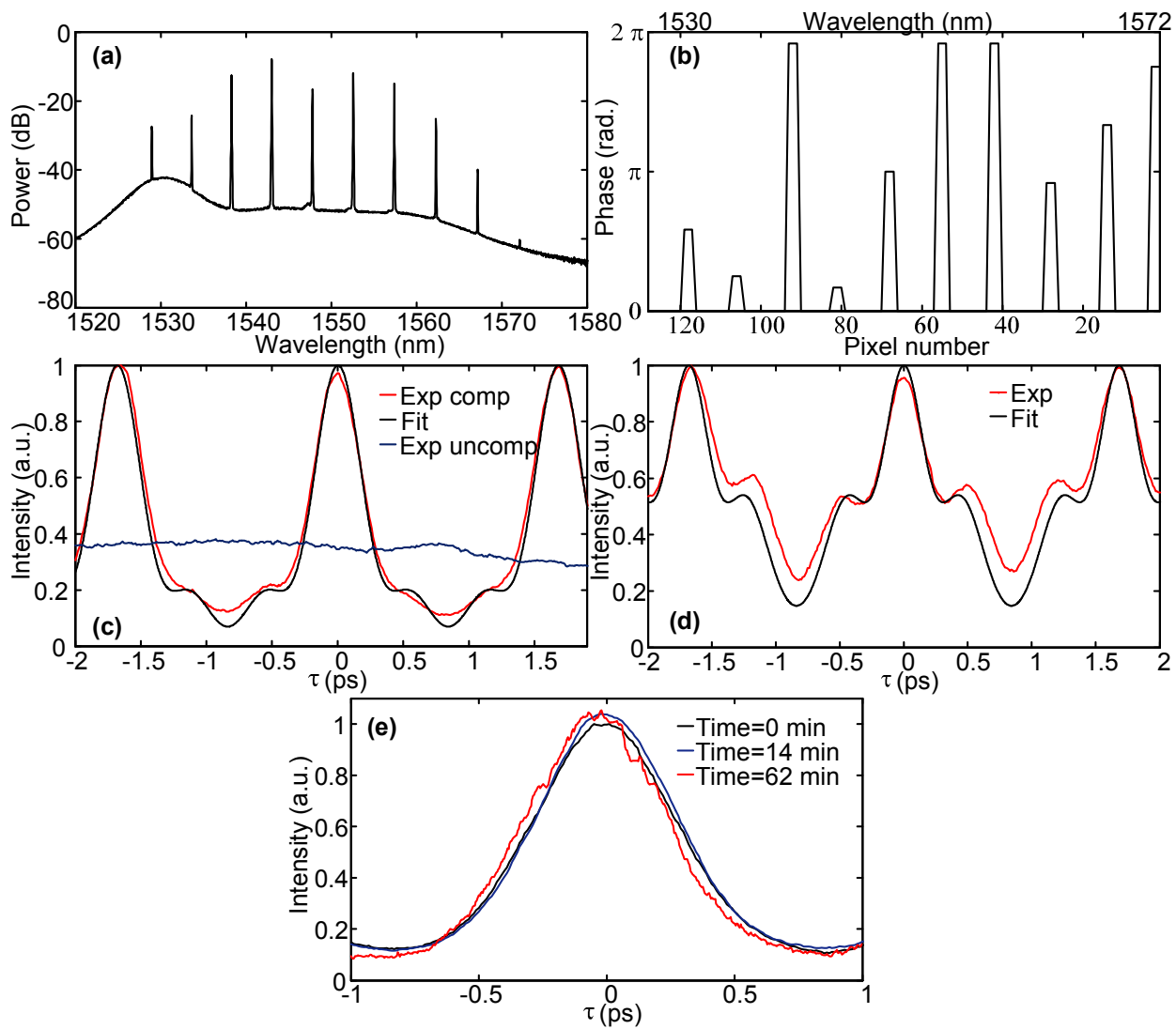